\documentclass{article}
\usepackage{fixltx2e}
\usepackage{authblk}
\usepackage[latin1]{inputenc}
\usepackage{amsmath,amsfonts,amssymb,bm}
\usepackage{graphicx}
\usepackage{cite}
\usepackage[tight,hang,small,bf]{subfigure}

\newcommand{\out}{{\rm out}}
\newcommand{\tloop}{t_{\rm loop}}

\newcommand{\philoop}{\phi_{\rm loop}}
\newcommand{\trep}{t_{\rm rep}}
\newcommand{\trans}{T_{12}}
\newcommand{\refls}{R_{12}}

\begin{document}

\title{True random numbers from amplified quantum vacuum}

\author{M.~Jofre$^{1,*}$,~M.~Curty$^{2}$,~F.~Steinlechner$^{1}$,~G.~Anzolin$^{1}$,~J.~P.~Torres$^{1,3}$,\\M.~W.~Mitchell$^{1,4}$~and V.~Pruneri$^{1,4}$}

\affil{\small{$^{1}$ ICFO-Institut de Ciencies Fotoniques, Castelldefels, E-08860 Barcelona, Spain.\\
$^{2}$ ETSI Telecomunicaci\'{o}n, Dept. Signal Theory and Communications, University of Vigo, E-36310 Vigo, Spain.\\
$^{3}$ Dept. Signal Theory and Communications, Universitat Polit\`{e}cnica de Catalunya, E-08034 Barcelona, Spain.\\
$^{4}$ ICREA-Instituci\'{o} Catalana de Recerca i Estudis Avan\c{c}ats, E-08010 Barcelona, Spain.}}

\affil{marc.jofre@icfo.es} 

\date{}

\maketitle

\begin{abstract}
Random numbers are essential for applications ranging from secure communications to numerical simulation and quantitative finance. Algorithms can rapidly produce pseudo-random outcomes, series of numbers that mimic most properties of true random numbers while quantum random number generators (QRNGs) exploit intrinsic quantum randomness to produce true random numbers. Single-photon QRNGs are conceptually simple but produce few random bits per detection. In contrast, vacuum fluctuations are a vast resource for QRNGs: they are broad-band and thus can encode many random bits per second. Direct recording of vacuum fluctuations is possible, but requires shot-noise-limited detectors, at the cost of bandwidth. We demonstrate efficient conversion of vacuum fluctuations to true random bits using optical amplification of vacuum and interferometry. Using commercially-available optical components we demonstrate a QRNG at a bit rate of $1.11$ Gbps. The proposed scheme has the potential to be extended to $10$ Gbps and even up to $100$ Gbps by taking advantage of high speed modulation sources and detectors for optical fiber telecommunication devices.
\end{abstract}

\section{Introduction}
The need for random numbers in research and technology was recognized very early \cite{Galton1890}, and has motivated electronic and photonic advances \cite{RAND1955,Kanai2009,Ribordy2007}. Random numbers support critical activities in advanced economies, including secure communications \cite{Gisin2002,Cerf2009,Ferguson2010}, numerical simulation \cite{Metropolis1949} and quantitative finance \cite{Banks2008}. For this reason, there has been intense effort to develop practical true random number generators, to replace existing pseudo-random methods. QRNGs employ a true source of randomness known to science, the randomness embedded into quantum physics. Recently, it has been shown that quantum physics also can be used to verify the randomness of entanglement-based generators \cite{Pironio2010,Colbeck2011}.

Examples of demonstrated QRNGs include two-path splitting of single photons \cite{Jennewein2000}, photon-number path entanglement \cite{Kwon2009}, time of generation or counting of photons \cite{Stipcevic2007,Bronner2009,Wayne2010,Fuerst2010,Wahl2011}, fluctuations of the vacuum state using homodyne detection techniques \cite{Gabriel2010,Symul2011} as well as interferometric schemes \cite{Uchida2008,Qi2010,Guo2010}.

Although any quantum measurement provides some randomness, a practical source must be simultaneously fast, inexpensive, and robust. For this purpose, fluctuations of the quantum vacuum are very attractive because the electric field amplitude is a continuous quantity, a single measurement can yield many true random bits. True vacuum is also perfectly white, uncorrelated, and broadband; the quantum field renews its random value arbitrarily quickly. Guaranteeing true vacuum is far from trivial, however; any scattered light will contribute a non-random component to the field measurement. Here we demonstrate extraction of random bits from vacuum using optical amplification. Homodyne detection based schemes \cite{Trifonov2007,Gabriel2010,Symul2011} guarantee that the signals originate in vacuum noise. Our method relies on the vacuum noise fluctuations as homodyne detection schemes, and at the same time achieves high bandwidth, because the requirement for shot-noise-limited detection is removed.

Relative to demonstrated methods for QRNG and achieved speeds, our proposed device is not only highly integrated, using commercially available components, but also has other advantages. In particular, the strong current modulation, well above and below threshold, ensures true randomness from vacuum. This active gain control allows a single device to have both a short coherence time, for rapid extraction of uncorrelated random bits, and a high signal level. In this way, standard photodiodes can be used. Furthermore, due to the high power of the signal pulses, the signal-to-noise ratio (SNR) is high. Hence, several random bits per detection event can be generated, limited by the classical noise of the measurement equipment. To our knowledge, it is the first time that our use of current gain modulation is used in QRNG.

\section{Device operation}
\begin{figure}[htbp]
\begin{center}
\subfigure[Electrical and optical pulse trains generated.]{\label{Fig:Laser_output_and_Laser_driver_output_time}\includegraphics[angle=0,width=0.5\textwidth]{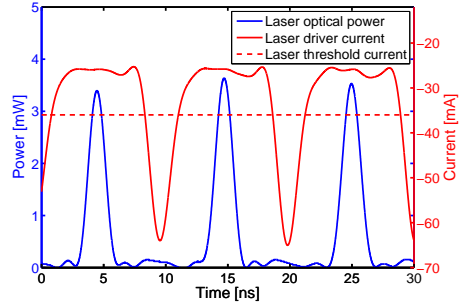}}
\subfigure[Device optical scheme.]{\label{Fig:Mach-Zehnder_Scheme}\includegraphics[angle=0,width=1\textwidth]{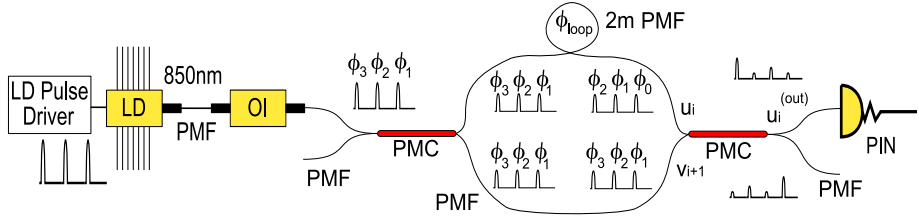}}
\caption{Unbalanced Mach-Zehnder interferometer. Due to the random phase of the different input pulses, the output signals acquire random amplitudes.  (a) Measured drive current (red, upper curve) and detected laser power (blue, lower curve), showing amplitude repeatability and clear pulse separation. (b) (LD Pulse Driver) denotes the electrical pulse generator to directly modulate the laser, (LD) laser diode, (OI) optical isolator, (PMF) polarization maintaining fiber, ($\phi_{0-3}$) optical phases of different consecutive pulses, (PMC) polarization maintaining coupler, ($\phi_{loop}$) phase introduced by the delay line and (PIN) fast photodiode.}
\label{Fig:Mach-Zehnder_Scheme_and_Laser_Pulses}
\end{center}
\end{figure}

We use a distributed feedback (DFB) laser diode (LD) as the oscillator, providing single-mode operation and high modulation bandwidth. The DFB LD is directly modulated at around $100$ MHz by a train of $\sim 1$ ns electrical pulses, as shown in  Fig. \ref{Fig:Laser_output_and_Laser_driver_output_time}. A polarization-maintaining, all-fiber unbalanced Mach-Zehnder interferometer (MZI) with a relative delay of $\tloop \approx {10}$ ns provides stable single-mode operation of the interferometer, as shown in Fig. \ref{Fig:Mach-Zehnder_Scheme}.

The LD is set with $25$ mA DC bias current, far below its threshold value of $36$ mA. Phase-randomized coherent optical pulses of $400$ ps time width and $3.5$ mW peak power are produced. A $30$ dB optical isolator (OI) is placed just after the LD to avoid back reflections into the oscillator cavity. Then, the linearly polarized optical pulses are split in power using a polarization maintaining coupler (PMC) with a fixed coupling ratio. In one of the output ports of the PMC, a $2$ m polarization maintaining fiber (PMF) patchcord is connected, which corresponds approximately to the equivalent length of the PRF. Both arms of the interferometer are connected to a second PMC where the interference between pulses takes place. The overall interferometer setup, at the output, has power coupling ratios of $T_{12}^{2}\approx 49.8$\% and $R_{12}^{2}\approx 40.3$\%, and polarization isolation of $23.98$ dB and $25.23$ dB for the two arms. At one of the output ports of the interferometer, a $150$ MHz photodiode is connected to collect the different interfering optical pulses which are processed by a fast oscilloscope. The oscilloscope is operated with a $200$ MHz bandwidth for the input channel, triggered by the system clock reference.

The path delay difference of the interferometer can be adjusted to temporally overlap subsequent pulses. On the one hand, the time delay between interfering pulses can be controlled by fine tunning the propagation properties of the long arm of the interferometer to change the parameter $\phi_{loop}$. For instance, by changing the temperature of the optical fiber one can produce a refractive index change and also thermal expansion of a wavelength for a $0.03$\textdegree C temperature change, corresponding to $4.25$ fs. Albeit, the time adjustment range achievable is limited compared to the pulse repetition period $\sim 10$ ns. On the other hand, the interferometer can be temperature stabilized to $0.01$\textdegree C to keep the parameter $\phi_{loop}$ and the PRF changed to increase or decrease the time between successive pulses. The time delay difference between both arms of the MZI is related to the PRF as $\Delta t = 1/$PRF, which allows an accurate and larger time adjustment range. The path delay difference of the interferometer was adjusted by setting the PRF at $97.6$ MHz.

\section{Laser physics analysis}
\begin{figure}[htbp]
\begin{center}
\subfigure[Loss mechanism scheme.]{\label{Fig:Quantum_GainSwitching_Scheme_Loss}\includegraphics[angle=0,width=0.25\textwidth]{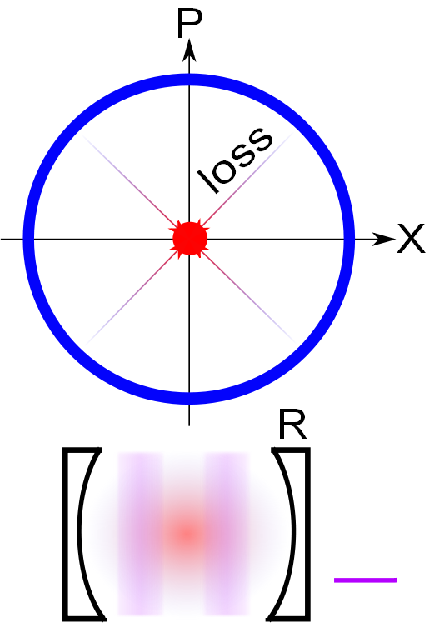}}
\subfigure[Gain mechanism scheme.]{\label{Fig:Quantum_GainSwitching_Scheme_Gain}\includegraphics[angle=0,width=0.25\textwidth]{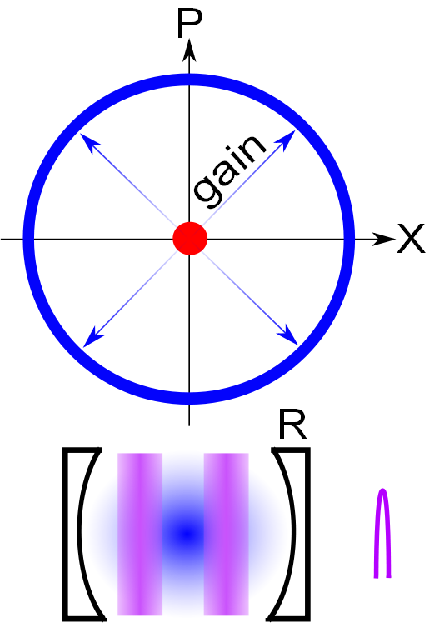}}
\caption{Generation of amplified vacuum within the laser cavity. (a) The LD is first taken below threshold, to attenuate the cavity field to a weak thermal state (in red), independent of its previous value (in blue). (b) The LD is then taken above threshold, so that phase-insensitive amplification brings the field amplitude $\left|\alpha\right|$ to a level fixed by saturation, while the phase retains the random thermal-state value.}
\label{Fig:Lasing_Mechanism}
\end{center}
\end{figure}

The method operates on the field within a single mode of a semiconductor diode laser. As shown in Fig. \ref{Fig:Lasing_Mechanism}, the laser is first operated far below threshold, producing simultaneously strong attenuation of the cavity field and input of amplified spontaneous emission (ASE). This attenuates to a negligible level any prior coherence, while the ASE, itself a product of vacuum fluctuations, contributes a masking field with a true random phase. The laser is then briefly taken above threshold, to rapidly amplify the cavity field to a macroscopic level. The amplification is electrically-pumped and thus phase-independent. Due to gain saturation, the resulting field has a predictable amplitude but a true random phase. The cycle is repeated, producing a stream of phase-randomized, nearly identical optical pulses. As shown in Fig. \ref{Fig:Mach-Zehnder_Scheme}, interference of subsequent pulses converts the phase randomness into a stream of pulses with random energies, which is directly detected and digitized.

During the attenuation phase, the cavity field is described by the Langevin equation:
\begin{equation}
\frac{d}{dt} a = -i \omega a- \frac{1}{2} \gamma a + \Gamma,
\end{equation}
where $a$ is the field operator for the mode, $\omega$ is its angular frequency, $\gamma$ is the (energy) decay rate and $\Gamma = {\gamma}^{1/2} a_{\rm res}  + \Gamma_{\rm ASE}$ is a noise operator, with $a_{\rm res}$ a reservoir mode.  The first term is from attenuation \cite{Collett1984}, and the second from ASE.  We can estimate $\gamma = \gamma_{\rm cav} + \gamma_{\rm mat}$ as follows: The cavity contribution is $\gamma_{\rm cav} = - c \ln(R)/(2 n L) = 5 \times 10^{10}$ s$^{-1}$, where $c$ is the speed of light in vacuum, $R=0.3$ is the out-coupler reflectivity, $n=3.6$ is the refractive index, and $L=300\mu$m is the cavity length. The material contribution $\gamma_{\rm mat} $ ranges from $c \alpha/n \approx 10^{11}$ s$^{-1}$ at zero current to $\gamma_{\rm mat} = -\gamma_{\rm cav}$ at threshold. Here $\alpha \approx 10^4 $cm$^{-1}$ is the intrinsic absorption of GaAs at $852$nm \cite{Sturge1962}. Interpolating, at $70$\% threshold current, we obtain $\gamma \approx 10^{11}$ s$^{-1}$, or about $400$ dB/ns. This renders completely negligible any prior coherence in the cavity, and the remaining field is an equilibrium between ASE and attenuation. The phase of this field is a true quantum random variable, its value determined by ASE which is driven by vacuum fluctuations. When the laser is taken above threshold, the equilibrated field is amplified, limited by gain depletion \cite{Suematsu2000}, to produce observed output powers of $P \approx 3.5$ mW or $1.5 \times 10^{7}$ photons/ns, with about $P/\gamma_{\rm cav} \approx 3 \times 10^5$ photons in the cavity. The amplification is phase-insensitive, and the phase of the cavity field remains truly random.

Considering the speed limits of this technique, we note that even at a modulation rate of $20$ GHz, i.e., an attenuation time of $\sim 0.25$ ns, the attenuation is $100$ dB. The field contribution remaining from the previous pulse is $3 \times 10^{-5}$ photons, or $\approx$ $15$ bits below the vacuum fluctuations. The physics of the process can thus support QRNG rates in excess of $100$ Gbps.

\section{Characterization of the coherence of the laser pulses}
\begin{figure}[htbp]
\begin{center}
\subfigure[Input and output statistics.]{\label{Fig:Int_PhaseRandomPulses_Statistics_InOut}\includegraphics[angle=0,totalheight=0.24\textheight]{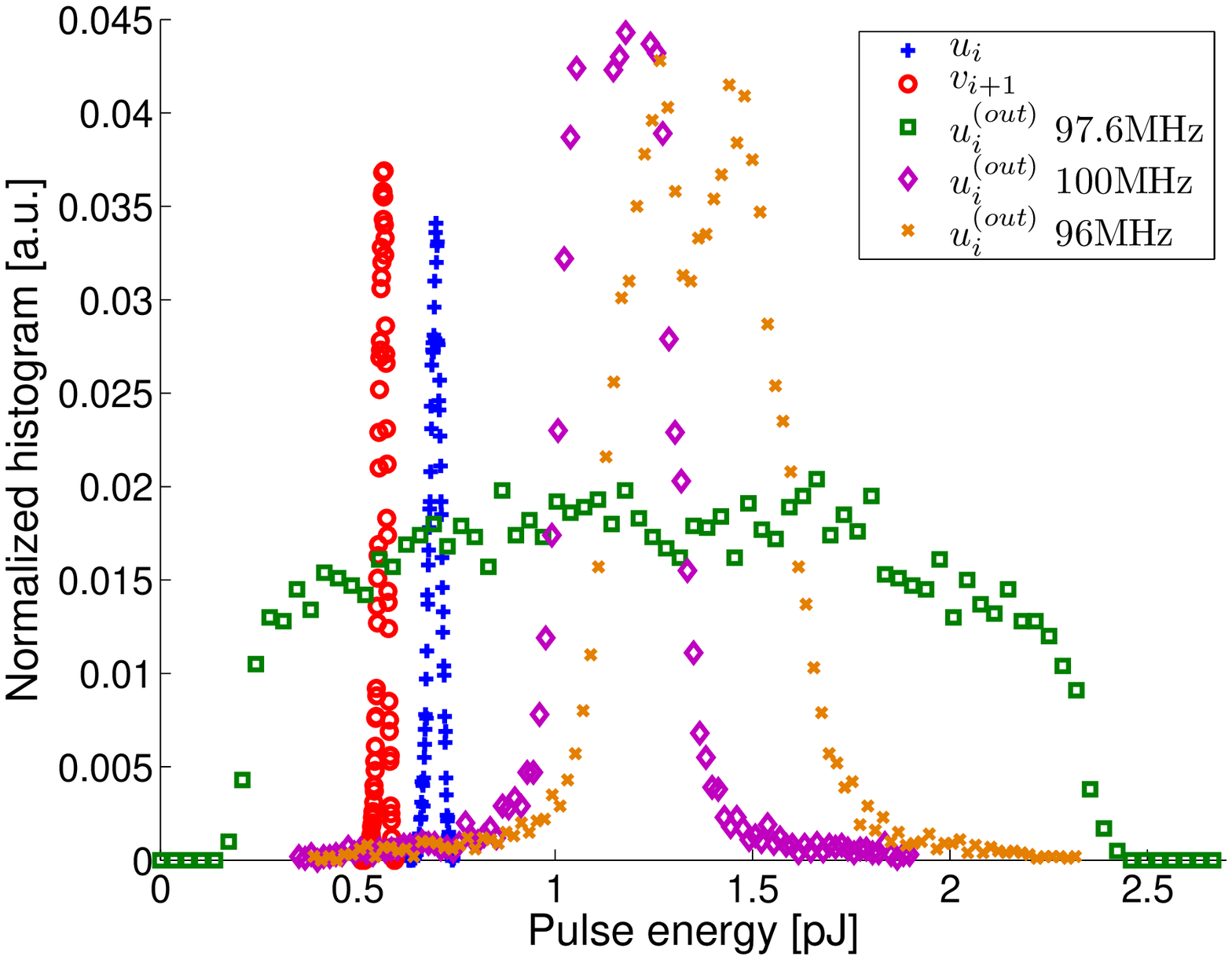}}
\subfigure[Statistics for different temperature settings.]{\label{Fig:Int_PhaseRandomPulses_Statistics_IntDifferentSettings}\includegraphics[angle=0,totalheight=0.24\textheight]{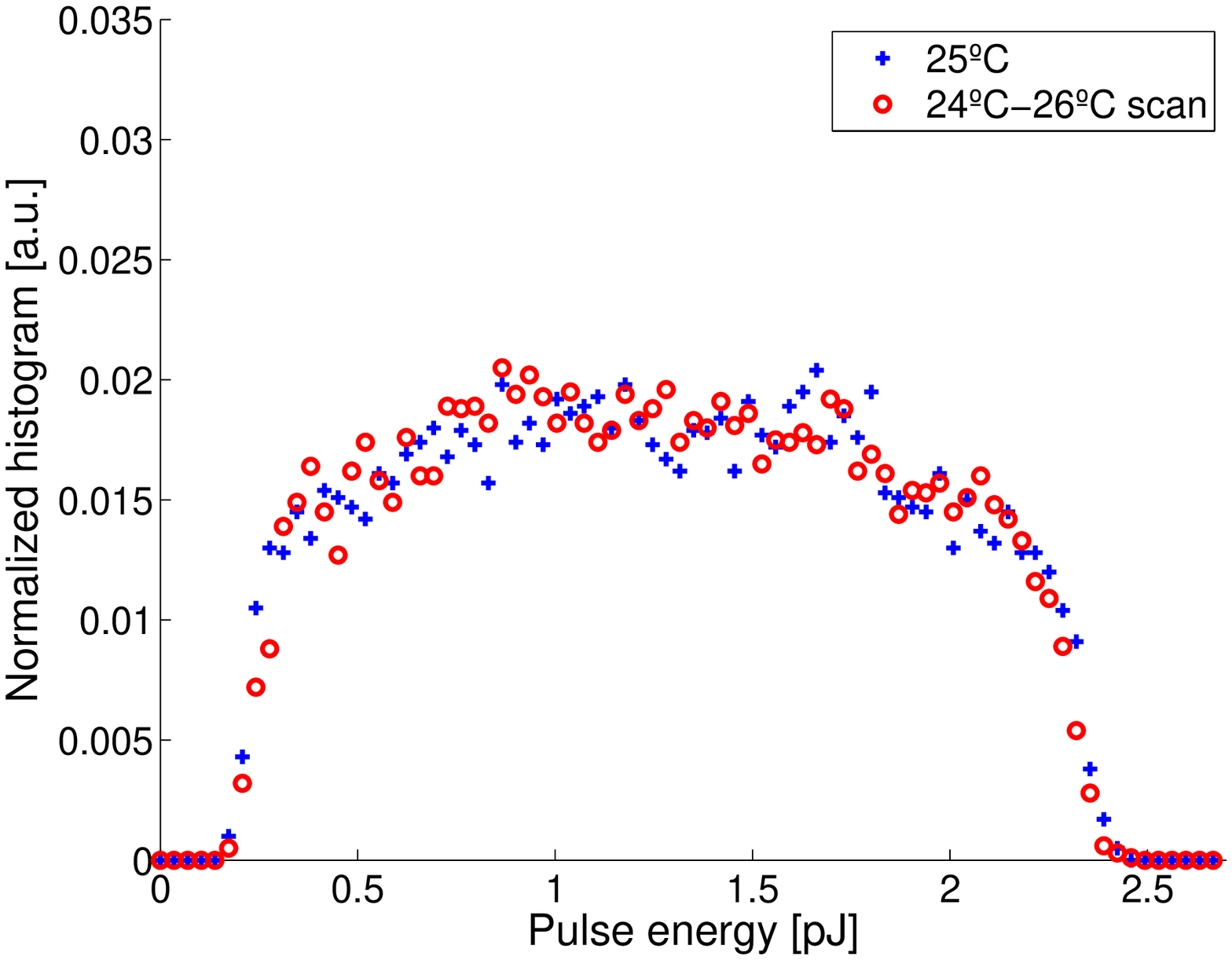}}
\caption{Inter-pulse coherence measured by output energy distributions. (a) Distributions for: individual pulse energies $u_i$,$v_{i+1}$, interfering pulse energies $u_i^{(\rm out)}$ under different PRF and hence different $\trep$. (b) Output pulse energy histogram for delay-loop temperatures of $25$ \textdegree C (fixed), and $24$ \textdegree C to $26$ \textdegree C (scanned). Loop phase has no observable effect on the distribution, indicating statistical independence of the pulses' phases.}
\label{Fig:Int_PhaseRandomPulses_Statistics}
\end{center}
\end{figure}

The interferometric setup allows us to determine the first order coherence properties of the laser pulses, described by the correlation functions
$G(\tau) \equiv \int{dt} \, \left< \right.  \hat{E}^{(-)}$ $(t)\hat{E}^{(+)}(t+\tau) \left. \right>$, or its normalized version $g(\tau) \equiv G(\tau) / G(0)$. Here $\hat{E}^{(\pm)}$ are the positive- and negative-frequency parts of the emitted field $\hat{E}$ and integrals are taken over the duration of the pulse. We expect the pulse energies $G(0)$ to be narrowly distributed, and $g(\trep)$ to have near-unit magnitude and random phase, where $\trep$ corresponds to the time between successive pulses given by the pulse repetition frequency (PRF), as subsequent pulses have very similar envelopes and random phases $\phi$. The interferometer output is $\hat{E}_{\rm out}\left(t\right) = \trans \hat{E}(t) + \refls \hat{E}(t+\tloop) $ where $\trans,\refls$ indicate combined transmission and reflection coefficients through the two beamsplitters. If we define the pulse energy in both arms of the interferometer as $u_i \equiv \refls^2  G(0)$ and $v_{i+1} \equiv \trans^2  G(0)$, the energy at the output port of the interferometer, $u_{i}^{(\out)} \equiv \int{dt} \, \left< \right. \hat{E}^{(-)}_{\rm out}(t)\hat{E}^{(+)}_{\rm out}(t)  \left. \right>_i $ is given by
\begin{eqnarray}\label{Eq:ReducedInterferometerForm}
 u_{i}^{(\out)} &=&  u_i + v_{i+1} + 2  |g(\tloop)|
 \sqrt{u_{i}v_{i+1}}\cos\left(\phi_{i}-\phi_{i+1}-\philoop\right) 
\end{eqnarray}
where $\philoop = \omega \tloop$ is the phase introduced by the delay loop. We measure the relevant statistics as follows (data shown in Fig. \ref{Fig:Int_PhaseRandomPulses_Statistics_InOut}): narrow distributions of $u_i$ and $v_{i+1}$ are directly observed by blocking one or the other path. Interference leads to a broadening of the observed distribution, with the broadest distribution corresponding to $\trep = \tloop$. From the width of the $u_i^{(\rm out)}$ distribution and the mean values of $u_i,v_{i+1}$, we can estimate the interference visibility $|g(\tloop)| \approx 90.22\%$. To demonstrate that the laser pulses are phase-uncorrelated, we collect statistics both for $\philoop$ fixed, and for $\philoop$ swept over several $\pi$, obtained by heating the fiber loop during acquisition. Results, shown in Fig. \ref{Fig:Int_PhaseRandomPulses_Statistics_IntDifferentSettings}, are statistically identical, indicating the absence of any phase relation between subsequent pulses.

\section{Statistical testing}
\begin{figure}[htbp]
\begin{center}
\subfigure[Normalized correlation of raw data.]{\label{Fig:Int_PhaseRandomPulses_Measurement_Corr}\includegraphics[angle=0,totalheight=0.24\textheight]{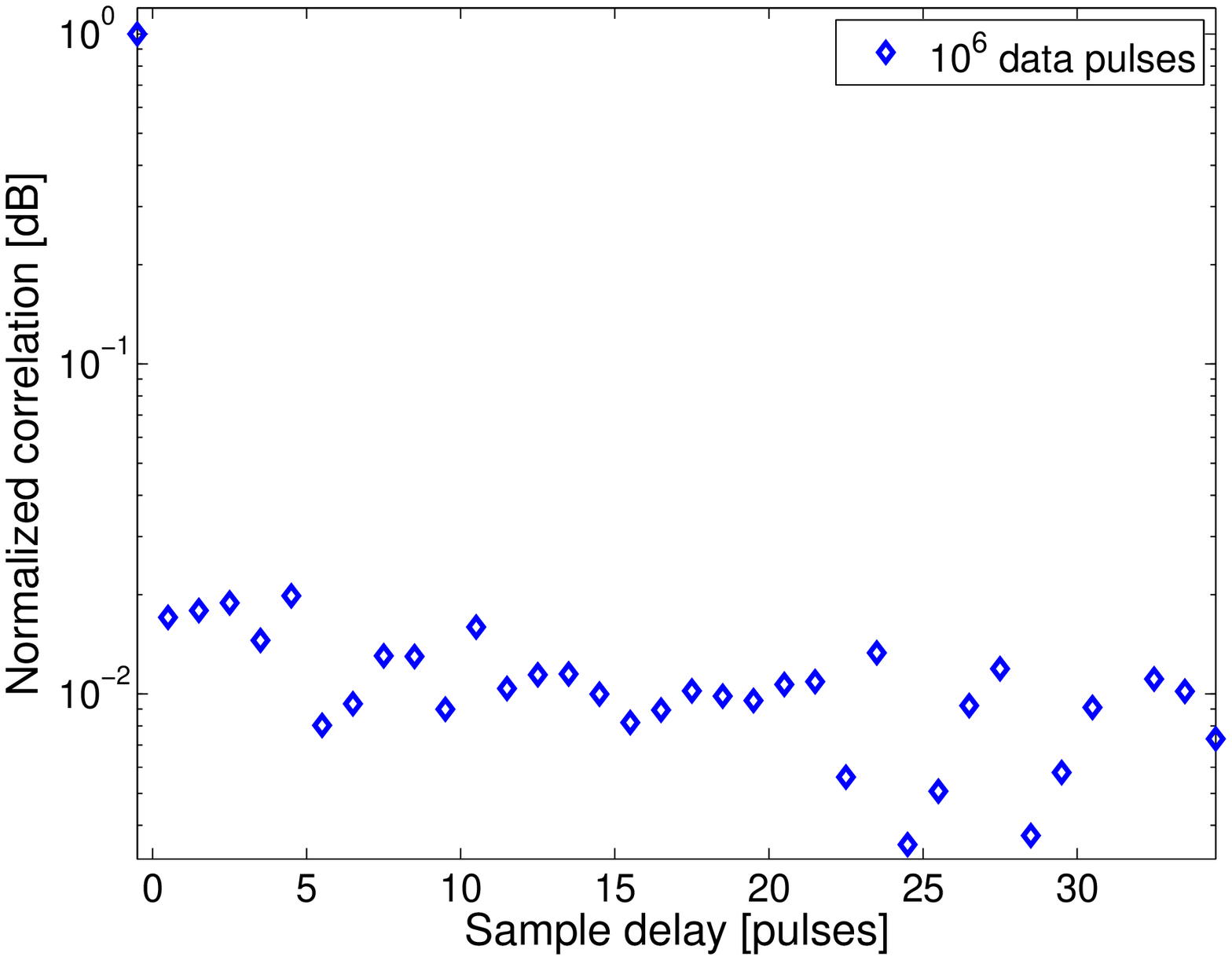}}
\subfigure[Total entropy of detected pulses and entropy of optical contribution .]{\label{Fig:Int_PhaseRandomPulses_Measurement_Entropy}\includegraphics[angle=0,totalheight=0.24\textheight]{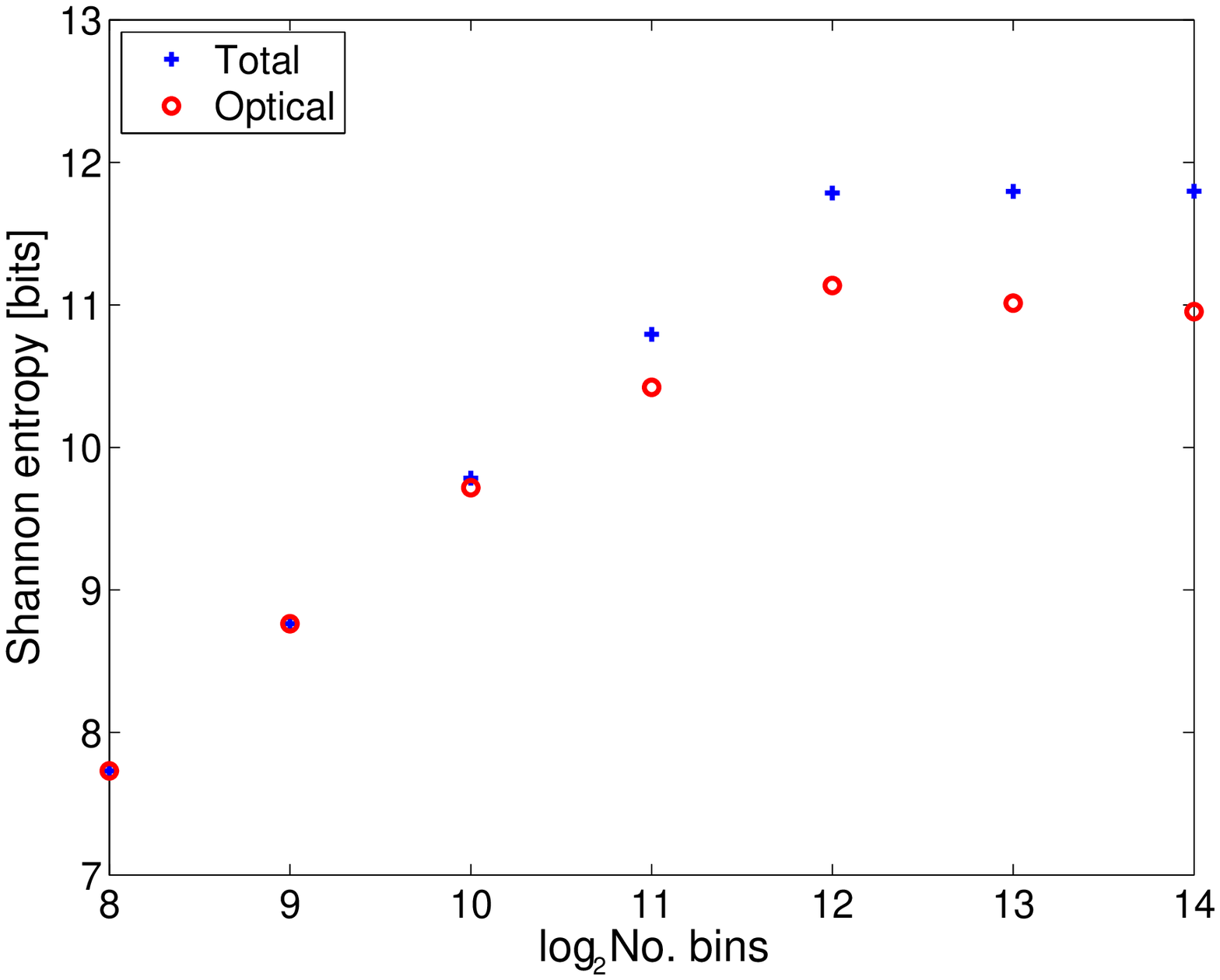}}
\caption{Measured correlation and entropy of acquired pulses. (a) Normalized correlation of successive samples as a function of sample delay of the raw data. The correlation data samples follows a delta-function like behavior indicating a random sequence. (b) Total entropy, calculated from the measured distribution shown in Fig. \ref{Fig:Int_PhaseRandomPulses_Statistics}. Distribution is divided into $2^b$ bins, from which the Shannon entropy is calculated. Optical contribution, up to $11.1$ bits per pulse, is found by subtracting entropy of the measured electronic noise.}
\label{Fig:Int_PhaseRandomPulses_StatisticsMeasurements}
\end{center}
\end{figure}

The output of the PIN photodiode was highpass filtered with a cutoff frequency of $40$ MHz and digitized using the waveform integration function of an oscilloscope with input bandwidth $200$ MHz, sampling speed of $2.5$ Gsps and a $12$-bit analog-to-digital converter (ADC). The $10$ ns time range setting, compliant with the PRF, and sampling speed of the oscilloscope permits to acquire $25$ samples over a pulse. The oscilloscope translates the multiple samples per pulse to a single measurement. The nearly uniform distribution of observed energies permits the use of equally-sized encoding bins, and facilitates calibration. Records of $10^{6}$ output pulses were collected in order to characterize the statistical correlations of the acquired raw data and to determine the number of extractable random bits per pulse. The normalized correlation of successive samples as a function of sample delay of the raw data is computed as the modulo-N circular auto-correlation for finite length sequences and it is normalized to the maximum, shown in Fig. \ref{Fig:Int_PhaseRandomPulses_Measurement_Corr}. The correlation of data samples follows a delta-function like behavior which indicates a random sequence with low impact of drifts in the system. The quantum random bit content of the recorded signal is determined as follows: The pulse distribution of Fig. \ref{Fig:Int_PhaseRandomPulses_Statistics} is divided into $2^{b}$ equally-sized bins and the Shannon entropy is calculated. As shown in Fig. \ref{Fig:Int_PhaseRandomPulses_Measurement_Entropy}, the entropy increases linearly with $b$, up to the value $b=12$, where it saturates to $11.8$ bits of entropy. The same procedure, applied to the detection noise, finds the classical noise entropy. Subtracting the noise entropy, the quantum optical noise contribution reaches a level of $11.1$ bits per pulse at $b=12$. Multiple samples per pulse achieves larger accuracy when used together with higher resolution ADC. This allows to better bound the contribution of the classical noise and thus permits to extract more true random bits per pulse.

The observed classical noise, however random it may appear, could in principle be the result of a completely predictable process. Indeed, randomness tests (described below) detect patterns in the recorded classical noise. To completely remove these patterns, we first note that the entropy of the classical noise places an upper bound on the information it can contain. We then remove this quantity of information, using cryptographic hash functions, from the combined quantum and classical noise \cite{Gabriel2010}. We use the Whirlpool hash function \cite{Barreto2010}; other standard randomness extractors could have also been employed \cite{Peres1992,Nisan1999}. These cryptographic functions mix the input data bits, increasing the theoretically secure entropy per bit at the cost of losing output bits. The reduction factor of the hash function applied to the collected raw bits is $1.08$. As a result, we obtain that the random bit generation rate of the current device accounts to $1.11$ Gbps.

We have performed all tests of randomness from TestU01 \cite{LEcuyer2007}. Considering the optical pulse data set, some test fail when applied to the raw data set, while they were successfully passed when applied to the hashed data set. Confirming that the hashing removes any remaining predictable behavior and increases the entropy per bit. Instead, the classical noise data set fails some tests both before and after hashing, using the same hashing factor.

\section{Conclusions}
In conclusion, we have demonstrated high-bandwidth extraction of random bits from quantum vacuum fluctuations using optical amplification. The use of strong attenuation followed by amplification guarantees that the signal originate from quantum noise, and provides macroscopic signals compatible with the highest bandwidth detection. With commercially-available components, we demonstrate over $1$ Gbps true random number generation. The QRNG device is low power consumption, robust, and can be easily automated allowing it to have a long operational lifetime. Consideration of the laser physics indicates that rates above $10$ Gbps and even $100$ Gbps are possible. The high random numbers generation rate extends the practical applications of our method to erode the dominance of currently used classical RNG choices. The method can be applied to high speed secure communication, to the gambling industry and to cryptography.

\section*{Acknowledgments}
The authors thank K. Tamaki, B. Qi, X. Ma and C. Wittmann for stimulating discussions. This work was carried out with the financial support of Xunta de Galicia (Spain) through grant INCITE08PXIB322257PR, and the Ministerio de Educaci\'on y Ciencia (Spain) through grants TEC2010-14832, FIS2007-60179, FIS2008-01051, FIS2010-14831 and Consolider Ingenio CSD2006-00019. This work is also supported by FONCICYT-94142.

\bibliographystyle{IEEEtran}

\end{document}